\documentclass[epj,final]{svjour}
\usepackage{graphics}
\usepackage{psfig}
\begin{document}
\title{Rolling as a ``continuing collision''}

\author{Nikolai V. Brilliantov\inst{1,2} \and Thorsten P\"oschel\inst{2}}

\institute{ \inst{1} Moscow State University, Physics Department, Moscow 
  119899, Russia\\
  \inst{2} Humboldt-Universit\"at, Institut f\"ur Physik, Invalidenstr. 
  110, D-10115 Berlin, Germany\\ http://summa.physik.hu-berlin.de/$\sim$kies/}
\date{Received: \today /  Revised version: }

\abstract{ We show that two basic mechanical processes, the collision
  of particles and rolling motion of a sphere on a plane, are
  intimately related. According to our recent findings, the
  restitution coefficient for colliding spherical particles
  $\epsilon$, which characterizes the energy loss upon collision, is
  directly related to the rolling friction coefficient $\mu_{\rm
    roll}$ for a viscous sphere on a hard plane. We quantify both
  coefficients in terms of material constants which allow to determine
  either of them provided the other is known.  This relation between
  the coefficients may give rise to a novel experimental technique to
  determine alternatively the coefficient of restitution or the
  coefficient of rolling friction.
}
\maketitle

Is the collision of two spheres related to rolling motion of a sphere
on a plane? Probably the answer will be ``No'', if the idealised
version of these two basic processes (i.e. absolutely elastic
collision and ideal rolling without any resistance) is assumed.
However, in reality neither ideal collision nor ideal rolling occur
since in both processes mechanical energy is lost according to
dissipative material deformation. Quantifying the losses a profound
similarity of these two processes has been revealed. It results in a
relation between quantities describing energy loss due to collision
and rolling and provides a novel view on rolling which may be
considered as a continuing collision.

It is well known since Newton's times that two balls colliding with
velocities $v_1$ and $v_2$ have smaller after-col\-lis\-io\-nal
velocities $v_1^{\prime}$ and $v_2^{\prime}$ as compared with the
perfectly elastic collision. The decrease of the particle velocity or
in other words the loss of mechanical energy is described by the
coefficient of restitution $\epsilon$. For a pair of identical
particles colliding on a line one has:
 
\begin{eqnarray}
\label{1}
v_1^{\prime}=v_1-\frac{1+\epsilon}{2}g \\
v_2^{\prime}=v_2+\frac{1+\epsilon}{2}g \nonumber 
\end{eqnarray} 
where $g=v_1-v_2$ is the initial relative velocity. The ideal elastic
collision is described by $\epsilon=1$, so that $v_1^{\prime}=v_2$ and
$v_2^{\prime}=v_1$. The relative velocity $g=v_1-v_2$ changes at the
collision as
\begin{equation}
\label{2}
g^{\prime}=-\epsilon g
\end{equation} 

We want to remark that for the more general case of a collision Eqs.
(\ref{1},\ref{2}) hold true but the pre- and after-collisional
velocities are vectors $\vec{v}_{1/2}$, $\vec{v}_{1/2}^\prime$ while
$\vec{g}$ and $\vec{g}^\prime$ are the pre- and after-collisional
relative normal velocities
\begin{eqnarray}
  \vec{g}&=&\left[ \left(\vec{v}_1-\vec{v}_2\right)\cdot \vec{e} \right]\,
  \vec{e}
 \\
  \vec{g}^\prime&=&\left[\left(\vec{v}^\prime_1-\vec{v}_2^\prime\right)
  \cdot \vec{e} \right]\,\vec{e} \nonumber  
\end{eqnarray}
where $\vec{e} =
(\vec{r}_1-\vec{r}_2)/\left|\vec{r}_1-\vec{r}_2\right|$ with
$\vec{r}_{1/2}$ being the particle positions at the instant of
collision.

The non-ideal character of the collision originates from the 
{\em dissipative} force 
\begin{equation}
\label{Disslaw}
F_{\rm diss} =\frac32 A \, \rho \,\sqrt{\xi} \dot{\xi}
\end{equation}
acting on the colliding spheres which has been derived recently
\cite{BSHP,KuwabaraKono,MorgadoOppenheim}.  Here, $\xi$ is the
time-dependent compression of the particles during the collision
\begin{equation}
\xi=2R-\left|\vec{r}_1-\vec{r}_2\right|
\end{equation}
with $R$ being the radius of the spheres, see Fig.1. The material
constant $\rho$ depends on the Young modulus $Y$ and Poisson ratio
$\nu$ of the particle material as
\begin{equation}
\rho = \frac{2Y}{3(1-\nu^2)}\sqrt{R/2}
\end{equation}
\begin{figure}[htbp]
\begin{minipage}{8.5cm}
\centerline{\psfig{file=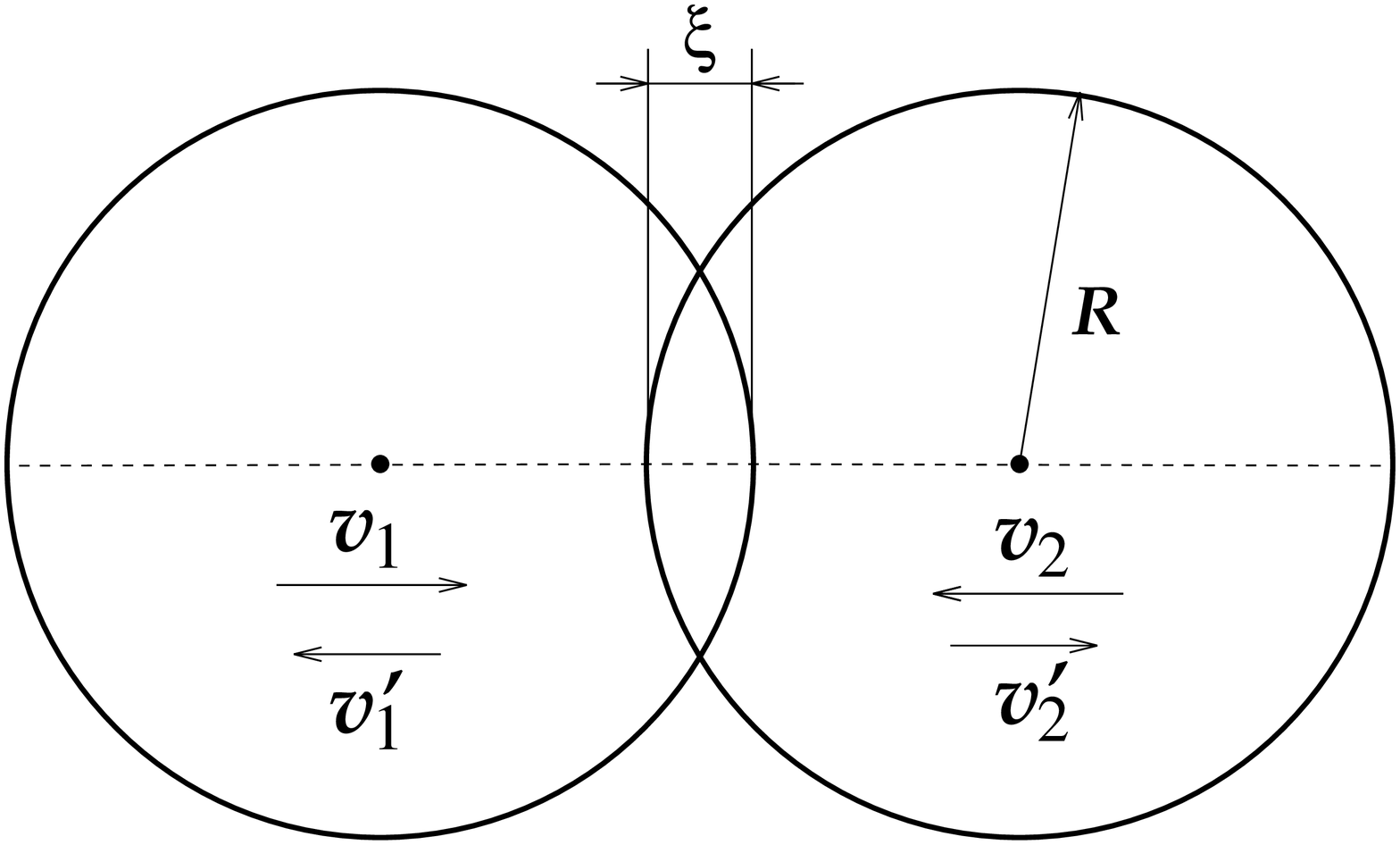,width=8cm}}
\vspace{1cm}
\centerline{\psfig{file=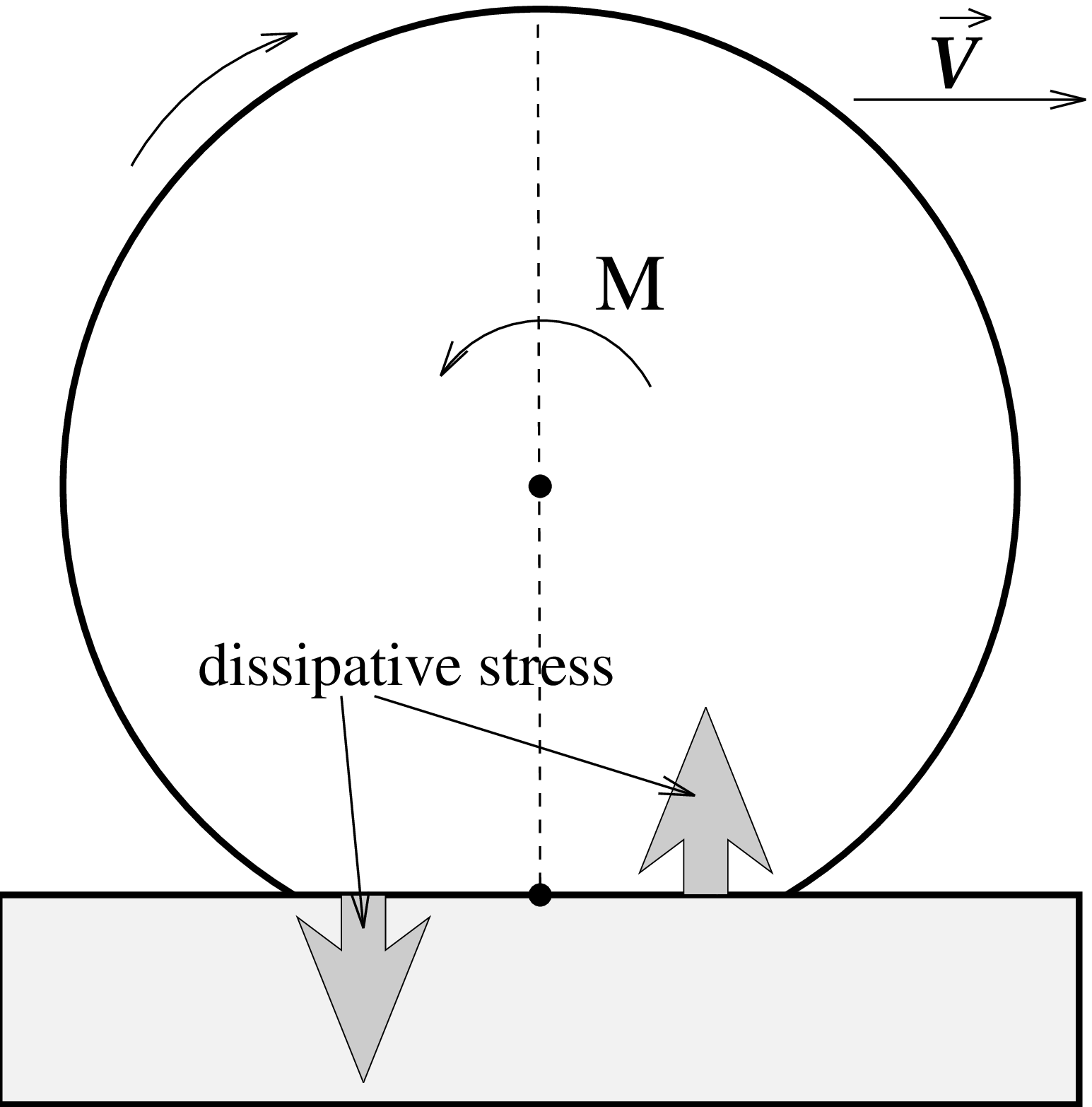,width=6cm}}
\vspace{0.5cm}
\caption{Sketches of two colliding spheres and of a rolling sphere. 
  Origination of the rolling friction moment due to the dissipative
  stress is shown.}
\label{fig:1}
\end{minipage}
\end{figure}

The constant $A$ is expressed in terms of the elastic and viscous
material constants \cite{BSHP}. The dissipative force always acts
against the relative velocity $\dot{\xi}$ of the particles, so that
elastic energy stored during the collision is not completely
reconverted into kinetic energy after the collision. In linear
approximation with respect to the dissipative parameter $A$ the
solution of the collision problem which accounts for the dissipative
force yields for the restitution coefficient~\cite{TomThor}
\begin{equation}
\label{eps}
1-\epsilon= C_1 \left(\frac{3A}{2} \right)
\left(\frac{2 \rho}{m} \right)^{2/5}g^{1/5} \pm \cdots
\end{equation}
with~\cite{TomThor,Rozaetal} 
\begin{equation}
C_1=\frac{ \Gamma(3/5)\sqrt{\pi}}{2^{1/5}5^{2/5} \Gamma(21/10)}=1.15344\,.   
\end{equation}
$\Gamma(x)$ is the Gamma-function and $m$ is the mass of the
(identical) particles. Hence, the restitution coefficient for
viscoelastic collisions depends sensitively on the normal component of
the relative velocity $g$ as described by Eq. (\ref{eps}).

Each collision of particles certainly terminates after some time if
attractive forces are excluded, but one can ask, whether some
``collision'' can proceed permanently.

In spite of being far from apparent one can show that the mechanics of
a rolling sphere on a hard plane is intrinsically similar to that of a
collision \cite{BTEurLet} provided the main part of the dissipation
originates from viscoelastic bulk deformations of the sphere
\cite{6,7,8}. Indeed, when a soft sphere rolls we notice the same
sequence of compression and subsequent decompression as in the
collision process. Thus, in rolling motion, which one can treat as a
``continuing'' collision, we observe a steady dissipation due to
incomplete retransformation of elastic energy during decompression.
This results in a rolling friction moment $M$, which acts against the
motion:
\begin{equation}
M=\mu_{\rm roll} F^{N}  
\label{Mdef}
\end{equation}
(see Fig.1). Here, $F^N$ is the normal force exerted by the plane
onto the sphere caused by the sphere's own weight. Calculations
performed for a soft sphere rolling on a hard (undeformable) plane
reveal that
\begin{equation}
\label{muroll}
\mu_{\rm roll} =AV\,,  
\end{equation}
where $V$ is the sphere's linear velocity and $A$ is exactly the {\em
  same} material constant as in the law of collision \cite{BTEurLet}.
The relation (\ref{muroll}) has been obtained for the case when the
deformation of the rolling sphere is small compared to its radius, the
velocity $V$ is much less then the speed of sound in the material, and
when the relaxation time of the rolling process, estimated as the
ratio of the sphere deformation and the velocity $V$ is much larger
then the dissipative relaxation times of the viscoelastic material
\cite{BTEurLet}.

Thus, 
\begin{equation}
\label{mueps}
\frac{1-\epsilon}{b\,  \left(\rho/m \right)^{2/5}g^{1/5}}=
\frac{\mu_{\rm roll}}{V}
\end{equation}
with 
\begin{equation}
b=3C_1/2^{3/5}=2.28296  
\end{equation}
relates the rolling friction coefficient $\mu_{\rm roll}$ to the
coefficient of normal restitution $\epsilon$ for identical particles
colliding with relative velocity $g$. 

The linear Eq.(\ref{mueps}) refers to the case of of small impact
velocities of the colliding spheres. This situation is the most
favourable for performing the experimental studies. One can, however,
apply a more general nonlinear relation which accounts for high-order
corrections of the dissipative parameter $A$.  This follows from the
high-order expansion for the restitution coefficient \cite{Rozaetal}:

\begin{eqnarray}
1-\epsilon&=&
C_1 \left(\frac{3A}{2} \right)
\left(\frac{2 \rho}{m} \right)^{2/5}g^{1/5} \nonumber \\
&-&C_2\left(\frac{3A}{2} \right)^2 \left(\frac{2 \rho}{m} 
\right)^{4/5}g^{2/5} \nonumber \\
&+&C_3\left(\frac{3A}{2}\right)^3 \left(\frac{2 \rho}{m} 
\right)^{6/5}g^{3/5} \nonumber \\
&-&C_4\left(\frac{3A}{2} \right)^4 \left(\frac{2 \rho}{m} 
\right)^{8/5}g^{4/5} \pm \cdots 
\end{eqnarray}
where \cite{Rozaetal}
\begin{eqnarray}
C_2&=&3/5\,C_1^2\nonumber\\
C_3&=&0.315119\,C_1^3\nonumber\\
C_4&=&0.161167\,C_1^4\nonumber  
\end{eqnarray}
Introducing then 
\begin{equation}
\label{Z}
Z \equiv b \left(\frac{\rho}{m} \right)^{2/5}g^{1/5}\,
\frac{\mu_{\rm roll}}{V}
\end{equation}
one can write the generalization of the linear Eq.(\ref{mueps}) as 
\begin{equation}
\label{epsZ}
\frac{1-\epsilon}{Z}=
1-0.6\,Z+0.315119\,Z^2-0.161167\,Z^3\pm \cdots
\end{equation}
 
Physically the relations (\ref{mueps}) and (\ref{epsZ}) are based on
the intrinsic mechanical similarity of the collision and rolling
processes. In practice Eq.(\ref{mueps}) [or (\ref{epsZ})] may be used
to determine either of coefficients $\epsilon$ or $\mu_{\rm roll}$
provided the other one is known.

The measurement of restitution coefficients for spheres is a
complicated experimental problem, in particular if one is interested
in collisions at very low impact velocity, e.g.~\cite{Bridges}. These
values are of great importance in several problems, e.g. for the
description of the kinetics of planetary ring material where the
particles typically collide with velocities of the range
$\left(10^{-2}\dots 10^{-3}\right)$ m/sec. The derived relation between
the coefficient of rolling friction and the coefficient of normal
restitution allows to determine the latter value by measuring the
resistance of a sphere against rolling on a hard plane which might be
experimentally less complicated.

\medskip 

\begin{acknowledgement}
  The authors want to thank Michel Louge and Thomas Schwager for
  helpful discussion
\end{acknowledgement}

\end{document}